\documentclass[twocolumn,aps,prl,superscriptaddress,amsmath,amssymb]{revtex4}
\usepackage[latin9]{inputenc}
\usepackage{amsmath}
\usepackage{amssymb}
\usepackage{graphicx}
\usepackage{esint}

\makeatletter
\@ifundefined{textcolor}{}
{%
 \definecolor{BLACK}{gray}{0}
 \definecolor{WHITE}{gray}{1}
 \definecolor{RED}{rgb}{1,0,0}
 \definecolor{GREEN}{rgb}{0,1,0}
 \definecolor{BLUE}{rgb}{0,0,1}
 \definecolor{CYAN}{cmyk}{1,0,0,0}
 \definecolor{MAGENTA}{cmyk}{0,1,0,0}
 \definecolor{YELLOW}{cmyk}{0,0,1,0}
}

\usepackage{float}\usepackage{makeidx}\usepackage{CJK}\usepackage{colordvi}\usepackage{color}

\makeatother

\begin{document}
\begin{CJK*}{GB}{gbsn}

\title{Dissociation of Feshbach molecules via spin-orbit coupling in ultracold
Fermi gases}

\author{Lianghui Huang}

\affiliation{State Key Laboratory of Quantum Optics and Quantum Optics Devices,
Institute of Opto-Electronics, Shanxi University, Taiyuan 030006,
P.R.China }

\author{Pengjun Wang}

\affiliation{State Key Laboratory of Quantum Optics and Quantum Optics Devices,
Institute of Opto-Electronics, Shanxi University, Taiyuan 030006,
P.R.China }

\author{Peng Peng}

\affiliation{State Key Laboratory of Quantum Optics and Quantum Optics Devices,
Institute of Opto-Electronics, Shanxi University, Taiyuan 030006,
P.R.China }

\author{Zengming Meng}

\affiliation{State Key Laboratory of Quantum Optics and Quantum Optics Devices,
Institute of Opto-Electronics, Shanxi University, Taiyuan 030006,
P.R.China }

\author{Liangchao Chen}

\affiliation{State Key Laboratory of Quantum Optics and Quantum Optics Devices,
Institute of Opto-Electronics, Shanxi University, Taiyuan 030006,
P.R.China }

\author{Peng Zhang}

\affiliation{Department of Physics, Renmin University of China, Beijing, 100872,
P.R.China}

\affiliation{Beijing Key Laboratory of Opto-electronic Functional Materials \&
Micro-nano Devices (Renmin University of China)}

\author{Jing Zhang$^{\dagger}$}

\affiliation{State Key Laboratory of Quantum Optics and Quantum Optics Devices,
Institute of Opto-Electronics, Shanxi University, Taiyuan 030006,
P.R.China }

\affiliation{Synergetic Innovation Center of Quantum Information and Quantum Physics,
University of Science and Technology of China, Hefei, Anhui 230026,
P. R. China}
\begin{abstract}
We study the dissociation of Feshbach molecules in ultracold Fermi
gases with spin-orbit (SO) coupling. Since SO coupling can induce
quantum transition between the Feshbach molecules and the fully
polarized Fermi gas, the Feshbach molecules can be dissociated by
the SO coupling. We experimentally realized this new type of
dissociation in ultracold gases of $^{{\rm 40}}$K atoms with SO
coupling created by Raman beams, and observed that the dissociation
rate is highly non-monotonic on both the positive and negative
Raman-detuning sides. Our results show that the dissociation of
Feshbach molecules can be controlled by new degrees of freedoms,
i.e., the SO-coupling intensity or the momenta of the Raman beams,
as well as the detuning of the Raman beams.
\end{abstract}
\maketitle
\end{CJK*}

\emph{Introduction. }Recently spin-orbit (SO) coupling has emerged
as one of the most exciting research directions in ultracold atom
physics. It plays a key role in a variety of systems and gives rise
to new phenomena ranging from topological insulators
\cite{Hasan,Qi,Kane,Bernevig} to Majorana fermions \cite{Maj}.
Ultracold atomic gases offer an unique platform for engineering
synthetic SO coupling due to the wide tunability of experimental
parameters. Many schemes of generating artificial Abelian and
non-Abelian gauge fields have been developed via atom-light
interaction \cite{Juzeliunas,Dalibard,Nature,Rew-Spiel}. An equal
combination of Rashba and Dresselhaus SO coupling was first realized
experimentally in a neutral atomic BEC by NIST group
\cite{spielman}, in which two atomic spin states are dressed by a
pair of counter-propagating laser beams with two-photon Raman
transition. Successively, several groups were realized
experimentally with the same scheme and studied the intriguing
properties of SO-coupled BEC
\cite{Shuai,Shuai1,Washington,Washington1,Purdue}.

In parallel, SO-coupled Fermi gases \cite{Rew-Zhang} have also
attracted a great deal of attentions, since SO coupling induces
coupling between spin-triplet and spin-singlet states and further
gives rise to non-trival topological order and Majorana fermions.
Lots of progresses has been made on the experimental exploration of
the SO-coupled Fermi gases. The SO-coupled non-interacting fermionic
$^{40}$K \cite{Wang2012} and $^{6}$Li \cite{Cheuk} atoms have been
investigated. Subsequently, SO-coupled Fermi gases were studied
experimentally near a Feshbach resonance. The SO-coupling-induced
shift of the binding energy of a Feshbach molecule has been observed
via both radio-frequency spectroscopy \cite{Fu2013a} and scattering
resonance induced by Raman beams \cite{Spielman-Fermi}. Furthermore,
it has been observed that SO coupling can coherently produce
$s$-wave Feshbach molecules from a fully polarized Fermi gas, and
induce a coherent oscillation between these two \cite{Fu2014}.

In this letter, we report that SO coupling can dissociate $s$-wave
Feshbach molecules formed by ultracold Fermi atoms in different
pseudo spin states. In our experiment we prepare Feshbach molecules
of ultracold $^{40}$K atoms in states $|F=9/2,m_{F}=-7/2\rangle$
($|$$\uparrow\rangle$) and $|F=9/2,m_{F}=-9/2\rangle$
($|$$\downarrow\rangle$), and then create an SO coupling by ramping
up two counter-propagating Raman beams, as in Refs.
\cite{spielman,Wang2012,Cheuk} (Fig. 1(a)). After this ramping
process, we measure the number of remaining Feshbach molecules as a
function of the Raman detuning of the spin-orbit coupling. We
observe that a significant loss of Feshbach molecules is induced by
the SO coupling. The maximum loss occurs on both the positive and
negative side of the Raman resonance. These observations are
consistent with our theoretical analysis. This analysis shows that
the loss effect is due to the SO-coupling-induced transition from
the Feshbach molecule state, in which the two atoms are in the
singlet pseudo spin state, to free-motion states of two atoms in the
polarized pseudo spin states
$|$$\uparrow\rangle_{1}$$|$$\uparrow\rangle_{2}$ or
$|$$\downarrow\rangle_{1}$$|$$\downarrow\rangle_{2}$. It is quite
different from radio-frequency (RF) beam induced dissociation of a
Feshbach molecule into two free atoms in the singlet state of
different hyperfine states $|$$\downarrow\rangle$ and
$|F=9/2,m_{F}=-5/2\rangle$ \cite{Regal2003,Chin2004}. Our work
demonstrates that SO coupling, as a momentum-dependent Zeeman field,
can entangle the two-atom internal state with the relative spatial
motion, and thus exhibits significantly different effect on Feshbach
molecule, comparing to a momentum independent Zeeman field.

\begin{figure}
\centerline{ \includegraphics[width=8.5cm]{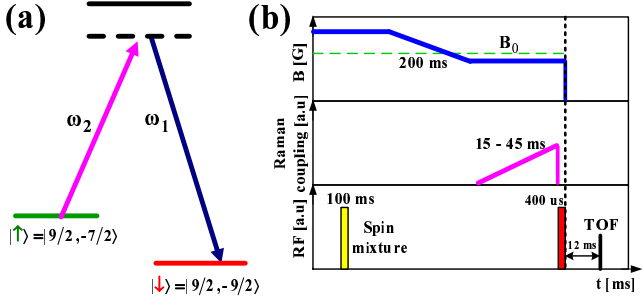} } 
\caption{(Color online). (a) Schematic diagram of SO coupling for $^{40}$K.
(b) The time sequence of the homogeneous bias magnetic field, the
Raman coupling and the RF field. Here $B_{0}=202.2$ G is the Feshbach
resonance point of atoms in $|F=9/2,m_{F}=-9/2\rangle$ and $|F=9/2,m_{F}=-7/2\rangle$.}

\label{Fig1}
\end{figure}

\emph{Experiment. } Our experimental setup for the preparation of
the Feshbach molecules and SO coupling in the ultracold Fermi gas
$^{40}$K has been described in detail in Refs.
\cite{Fu2013a,Fu2014}. We prepare an equal mixture of
$2\times10^{6}$ ultracold $^{40}$K atoms in the pseudospin states
$|$$\uparrow\rangle$ and $|$$\downarrow\rangle$, and then
adiabatically sweep the magnetic field across the Feshbach resonance
point $B_{0}=202.20$G (Fig. 1(b)). As a result of this adiabatic
sweeping, many pairs of atoms in state
$|$$\uparrow\rangle_{1}$$|$$\downarrow\rangle_{2}$ are converted
into the s-wave Feshbach molecules. The binding energy of the
molecules is determined by the final magnetic field strength in the
sweeping process. Subsequently, we apply the SO coupling by
switching on a pair of counter-propagating Raman laser beams
\cite{spielman,Wang2012,Cheuk}, which effectively couple the states
$|$$\uparrow\rangle$ and $|$$\downarrow\rangle$ (Fig. 1(a)). The
momentum transfer in the Raman process is $2k_{{\rm
r}}\equiv4\pi\hbar/\lambda$, where $\lambda=772.4$ nm is the
wavelength of the Raman beams. In our system the two-photon detuning
is defined as $\eta=\hbar(\omega_{1}-\omega_{2}-\omega_{Z})$, where
$\omega_{1,2}$ are the frequencies of the two Raman beams (Fig.
1(a)), and $\hbar\omega_{Z}$ is the Zeeman splitting between states
$|$$\uparrow\rangle$ and $|$$\downarrow\rangle$.

In each experiment, we fix the value of $\eta$ and ramp the
intensity of the Raman coupling from zero to a maximum value, and
then switch off the Raman beams (Fig. 1(b)). When the Raman beams
are switched off, we measure the number of remaining Feshbach
molecules in the trap with the approach in Refs.
\cite{Fu2013a,Fu2014}. We apply an RF pulse with duration about 400
$\mu s$ to dissociate these molecules into free atoms in states
$|$$\downarrow\rangle$ and $|F=9/2,m_{F}=-5/2\rangle$, and then
measure the number $N_{j}$ ($j=-7/2,\ -5/2$) of atoms in
$|F=9/2,m_{F}=j\rangle$ via time-of-flight technique. The number of
remaining molecules is known as $N_{{\rm rem}}=N_{-5/2}$, while
$N_{{\rm tot}}=N_{-5/2}+N_{-7/2}$ is half the number of all the
$^{{\rm 40}}$K atoms in our system.

\begin{figure}
\centerline{ \includegraphics[width=9cm]{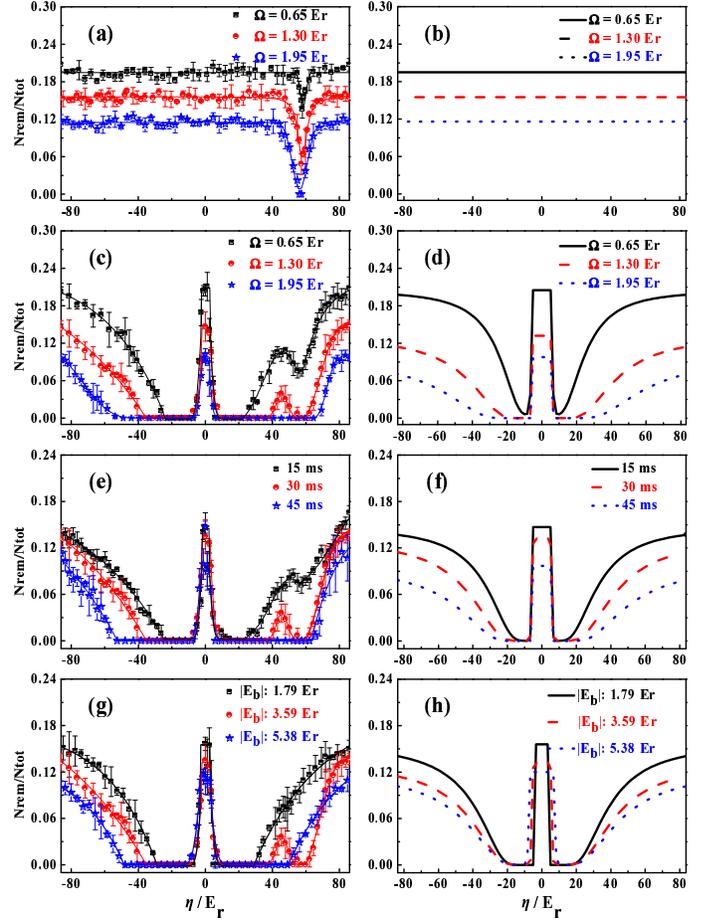} } 
\caption{(Color online). The ratio $N_{{\rm rem}}/N_{{\rm tot}}$
between the number $N_{{\rm rem}}$ of remaining Feshbach molecules
and half the number $N_{{\rm tot}}$ of all atoms. Here we illustrate
experimental (a, c, e, g) and theoretical (b, d, f, h) results as
functions of Raman detuning $\eta$. \textbf{(a)} and \textbf{(b)}:
results without SO coupling (the two Raman beams propagating along
the same direction). \textbf{(c)} and \textbf{(d)}: results with
binding energy of Feshbach molecule $|E_{b}|=3.59E_{r}$, ramping
time $T=30$ ms, and different values of final intensity $\Omega$ of
Raman coupling. \textbf{(e)} and \textbf{(f)}: results with
$|E_{b}|=3.59E_{r}$, $\Omega=1.30E_{r}$, and different $T$.
\textbf{(g)} and \textbf{(h)}: results with $\Omega=1.30E_{r}$,
$T=30$ ms and different $|E_{b}|$.}

\label{Fig2}
\end{figure}


In Figs. 2 (c, e, g), we illustrate the ratio $N_{{\rm rem}}/N_{{\rm
tot}}$ between the number of remaining molecules and half the number
of all the atoms, as a function of two-photon detuning $\eta$. It is
clearly shown that in some parameter regions this ratio approaches
zero. Therefore, in these regions most of the molecules are
dissociated. For comparison, we also do measurements (Fig. 2(a)) in
the system where the two Raman beams propagate along the same
direction (the other parameters are same as the Fig. 2(c)). In this
case, SO coupling cannot be created by the Raman beams, and we find
that the ratio $N_{{\rm rem}}/N_{{\rm tot}}$ is large and does not
change with $\eta$. Note that the narrow peak in the blue Raman
detuning of about 58 $E_{r}$ is due to the bound-to-bound (Feshbach
molecular state to deeply bound molecular states) transitions with
the Raman lasers \cite{Huang2014}. Therefore, in our system the
dissociation of the Feshbach molecules is induced by the SO
coupling. Below we will give a theoretical explanation for this
SO-coupling-induced dissociation. We show that in the presence of SO
coupling, the Raman beams can induce a transition from the Feshbach
molecule state to the free-motion states of two atoms in
$|$$\uparrow\rangle_{1}$$|$$\uparrow\rangle_{2}$ or
$|$$\downarrow\rangle_{1}$$|$$\downarrow\rangle_{2}$. As a result,
the molecules can be dissociated. Nevertheless, when there is no SO
coupling, this transition is forbidden by the symmetry of the
system.

Figures 2 (c, e, g) also show that, when the two-photon detuning
$\eta$ is zero, the ratio $N_{{\rm rem}}/N_{{\rm tot}}$ remains
unchanged compared with the case in which the SO coupling is not
applied. Therefore, in that case there is no dissociation effect.
Nevertheless, when $|\eta|$ is increased to a sufficient detuning,
$N_{{\rm rem}}/N_{{\rm tot}}$ rapidly decreases, which corresponds
to the steep threshold behavior of the dissociation process. In some
regions with finite $|\eta|$, $N_{{\rm rem}}/N_{{\rm tot}}$ is
negligible. This implies that in these regions the dissociation
effect is very strong and saturated. This phenomenon is due to the
energy conservation in the Raman-beam-induced transition. As shown
in the theoretical analysis below, when $\eta=0$, the Feshbach
molecule state lies in the lowest energy state of the system and all
the polarized states are energetically off-resonant with the
Feshbach molecule state. Thus there is no dissociation effect. When
the detuning $\eta$ takes a sufficient positive value, the Feshbach
molecule state becomes resonant with the lower free-motion states in
$|$$\uparrow\rangle_{1}$$|$$\uparrow\rangle_{2}$. As a result, the
transitions from the Feshbach molecule state to these states can
take place and are strong if $\eta$ is not too large. We thus can
observe a significant dissociation effect and a steep threshold
behavior. Similarly, when $\eta$ reaches a sufficient negative, the
Feshbach molecule state becomes resonant with polarized states in
$|$$\downarrow\rangle_{1}$$|$$\downarrow\rangle_{2}$, and thus
dissociation can take place.

Furthermore, as shown in Figs. (c, e, g), in the regions where
$|\eta|$ is extremely large, the ratio $N_{{\rm rem}}/N_{{\rm tot}}$
gradually increases with $|\eta|$. Therefore, in these regions the
dissociation effect becomes weak again. According to our theoretical
analysis, this is because that the Feshbach molecule state is
resonant with free-motion states with high momentum when $|\eta|$ is
very large. As a result, the matrix element of the Hamiltonian
between these two states (Frank-Condon factor) becomes small, and
thus the transition rate from the molecule state to the polarized
states is decreased.

We also investigate the dependence of the dissociation effect on
other physical parameters. In Fig. 2(c) we illustrate $N_{{\rm
rem}}/N_{{\rm tot}}$ measured with the binding energy of the
Feshbach molecule $|E_{b}|=3.59E_{r}$, ramping time $T=30$ ms, and
final intensity of Raman coupling in the ramping process
$\Omega=0.65E_{r}$, $1.30E_{r}$ and $1.95E_{r}$, where
$E_{r}=k_{{\rm r}}^{2}/(2m)=\hbar\times52.52$ kHz is the recoil
energy of the Raman beams. Here $m$ is the single-atom mass. In Fig.
2(e) we show the values of $N_{{\rm rem}}/N_{{\rm tot}}$ for
$|E_{b}|=3.59E_{r}$, $\Omega=1.30E_{r}$, and $T=15$ ms, $30$ ms and
$45$ ms. Our measurements show that the dissociation effect is
strong under the condition of long ramping time and high final
intensity of Raman coupling. In Fig. 2(g) we illustrate $N_{{\rm
rem}}/N_{{\rm tot}}$ for $T=30$ ms, $\Omega=1.30E_{r}$, and
$|E_{b}|=1.79E_{r}$, $3.59E_{r}$ and $5.38E_{r}$. We find that the
dissociation effect increases with the binding energy $|E_{b}|$ of
the Feshbach molecule (Note that the narrow peak for the
bound-to-bound transitions with the Raman lasers is shifted when the
magnetic field (binding energy) is changed \cite{Huang2014}). This
phenomenon can possibly be explained with the following analysis.
When $|E_{b}|$ becomes larger, the Feshbach molecule state has a a
broader momentum distribution. As a result, there are more
free-motion states which have large overlap (Frank-Condon factor)
with the molecule state.

In the following we present a detailed theoretical analysis for our
experiment. We theoretically calculate the ratio $N_{{\rm
rem}}/N_{{\rm tot}}$ with the same parameters as our experiments,
which agree well with the experimental measurements.

\emph{Theoretical analysis.} Our experimental results can be
qualitatively explained with a simple $2$-body analysis. For
convenience, here we discuss our problem in the co-moving frame
which is related to the original frame via a spin-dependent unitary
transformation ${\cal
U}=e^{-ik_{0}(x_{1}\sigma_{z}^{(1)}+x_{2}\sigma_{z}^{(2)})}$, with
$\sigma_{z}^{(i)}=$$|$$\uparrow\rangle_{i}\langle\uparrow$$|-|$$\downarrow\rangle_{i}\langle\downarrow$$|$
and $k_{0}=k_{{\rm r}}\sin\frac{\theta}{2}$, where $\theta$ is the
angle between two Raman beams. In this co-moving frame, the
Hamiltonian of the two atoms is $H=H_{1}+H_{2}$, with ($\hbar=m=1$)
\begin{eqnarray}
H_{1} & = & \sum_{i=1,2}\left[\frac{1}{2}\left({\bf p}^{(i)}+k_{0}\sigma_{z}^{(i)}{\bf e}_{x}\right)^{2}-\frac{\eta}{2}\sigma_{z}^{(i)}\right]+V,\label{hf-1}\\
H_{2} & = & \frac{\Omega}{2}\left(\sigma_{x}^{(1)}+\sigma_{x}^{(2)}\right).\label{hi-1}
\end{eqnarray}
Here ${\bf p}^{(i)}$ $(i=1,2)$ is the momentum of atom $i$, ${\bf
e}_{x}$ is the unit vector along the $x$-direction,
$\sigma_{x}^{(i)}=|$$\uparrow\rangle_{i}\langle\downarrow$$|+|$$\downarrow\rangle_{i}\langle\uparrow$$|$,
$\Omega$ is the Raman-coupling strength. In Eq. (\ref{hf-1}), $V$ is
the inter-atom interaction operator in the co-moving frame. In the
low-energy case, we only consider the interaction between fermionic
atoms in different pseudo spin states. Based on this model, we can
explain our experimental results.

\textit{(1) Finite $k_{0}$ is necessary for the dissociation of
Feshbach molecule.} According to Eqs. (\ref{hf-1}) and (\ref{hi-1}),
when the Raman beams are turned on, the pseudospin-dependent part of
the total Hamiltonian $H$ of the two atoms can be written as ${\bf
h}({\bf p}^{(1)}){\bf \sigma}_{1}+{\bf h}({\bf p}^{(2)}){\bf
\sigma}_{2}$, where ${\bf h}({\bf p}^{(i)})=\frac{\Omega}{2}{\bf
e}_{x}+p_{x}^{(i)}k_{0}{\bf e}_{z}$ is the effective Zeeman field
experienced by atom $i$. In the system where the two Raman beams
propagate along the same direction, we have $k_{0}$$=$$0$ , and
there is no synthetic SO coupling. As a result, the effective field
is momentum-independent. Thus, when the Raman beams are applied, the
pseudospins of the two atoms rotate along the same axis.
Furthermore, the two atoms in the Feshbach molecule are in the
singlet state $|S\rangle$, which cannot be changed by such rotation.
Therefore, the Raman beams cannot dissociate the Feshbach molecule.
On the other hand, when the two Raman beams propagate along
different directions, we have $k_{0}\neq0$
 , and the synthetic SO coupling is induced by the Raman beams.
In this case the two atoms with different momentum ${\bf k}_{1}$ and
${\bf k}_{2}$ can experience different effective fields ${\bf
h}({\bf k}_{1})$ and ${\bf h}({\bf k}_{2})$. Thus, the Raman beams
can rotate the pseudospins of the two atoms along different axis.
Therefore, although the two atoms in the Feshbach molecule are
polarized along opposite directions, when the Raman beams are turned
on, they have some probability to evolve to the parallel-polarized
state where the pseudospins are along the same direction. As a
result, the Feshbach molecule can be dissociated by the Raman beams.

This result can also be understood with the following detailed
analysis. According to Eqs. (\ref{hf-1}) and (\ref{hi-1}), before
the Raman beams are applied, we have $\Omega=0$ and thus $H_{2}=0$.
Therefore, the two-atom Hamiltonian in the co-moving frame is
$H_{1}$. The atoms are prepared in the Feshbach molecule state. In
the original frame, this state is $|\Phi_{b}\rangle=\int d{\bf
r}_{1}d{\bf r}_{2}\phi_{b}({\bf r})|{\bf r}_{1}\rangle_{1}|{\bf
r}_{2}\rangle_{2}|S\rangle,$ where
$|S\rangle=(|$$\uparrow\rangle_{1}|$$\downarrow\rangle_{2}-|$$\downarrow\rangle_{1}|$$\uparrow\rangle_{2})/\sqrt{2}$
is singlet state, $|{\bf r}_{i}\rangle_{i}$ is the eigen-state of
the position of the $i$th atom, ${\bf r}={\bf r}_{1}-{\bf r}_{2}$
and $\phi_{b}({\bf r})=e^{-r/a}/\sqrt{2a\pi}$, with $a$ the
scattering length between atoms in states $|$$\uparrow\rangle$ and
$|$$\downarrow\rangle$. Therefore, in the co-moving frame the
Feshbach molecule state is $|\Phi_{b}^{(C)}\rangle\equiv{\cal
U}|\Phi_{b}\rangle$. It is an eigen-state of $H_{1}$, with
eigen-energy $E_{b}=-a^{-2}$.

In the system with $k_{0}=0$, we have ${\cal U}=1$ and thus $|\Phi_{b}^{(C)}\rangle=|\Phi_{b}\rangle\propto|S\rangle$.
When the Raman beams are turned on, the atom-laser interaction is
described by the Hamiltonian $H_{2}$ in Eq. (\ref{hi-1}). Nevertheless,
since $H_{2}|S\rangle=0$, the Raman beams cannot induce quantum transition
from $|\Phi_{b}^{(C)}\rangle$ to other states. Therefore, the Raman
beams cannot dissociate the Feshbach molecule.

When $k_{0}\neq0$, we have ${\cal U}\neq1$. In this case the
Feshbach molecule state in the co-moving frame is
$|\Phi_{b}^{(C)}\rangle={\cal
U}|\Phi_{b}\rangle=|\phi_{+}\rangle|S\rangle+|\phi_{-}\rangle|T\rangle$,
where
$|T\rangle=(|$$\uparrow\rangle_{1}|$$\downarrow\rangle_{2}+|$$\downarrow\rangle_{1}|$$\uparrow\rangle_{2})/\sqrt{2}$
is triplet state, and $|\phi_{\pm}\rangle=\frac{1}{2}\int d{\bf
r}_{1}d{\bf r}_{2}\left[\phi_{b}({\bf r})e^{-ik_{0}{\bf
e}_{x}\cdot{\bf r}}\pm\phi_{b}({\bf r})e^{ik_{0}{\bf e}_{x}\cdot{\bf
r}}\right]|{\bf r}_{1}\rangle_{1}|{\bf r}_{2}\rangle_{2}.$ It is
clear that $H_{2}|\Phi_{b}^{(C)}\rangle\propto
H_{2}|T\rangle\propto(|$$\uparrow\rangle_{1}|$$\uparrow\rangle_{2}+|$$\downarrow\rangle_{1}|$$\downarrow\rangle_{2})$.
Therefore, when the Raman beams are turned on, the atom-laser
interaction $H_{2}$ can induce quantum transition from the Feshbach
molecule state $|\Phi_{b}^{(C)}\rangle$ to other eigen-states of
$H_{1}$, and thus dissociate the Feshbach molecules.

\begin{figure}
\centerline{ \includegraphics[width=8.5cm]{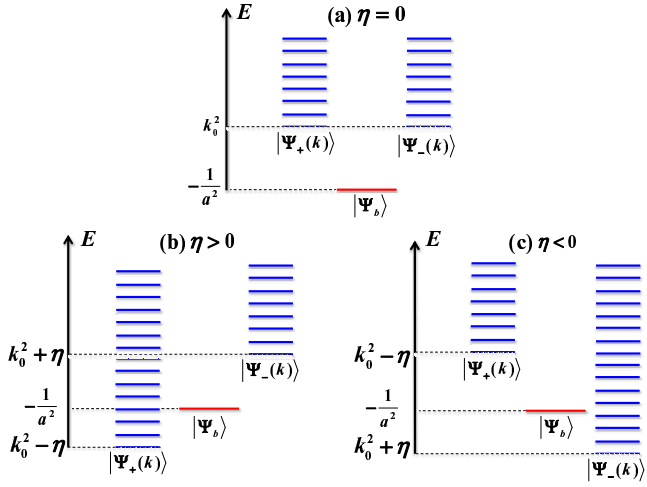} } 
\caption{(Color online) The energy spectrums of eigens-states of
$H_{F}$ with two-photon detuning $\eta=0$ (a), $\eta>0$ (b) and
$\eta<0$ (c).}

\label{Fig3}
\end{figure}

\textit{(2) Non-monotonic dependence of dissociation on detuning
$\eta$.} Now we consider the dependence of the dissociation effect
on the two-photon detuning $\eta$. As shown above, the dissociation
is due to Raman-beams-induced transition from the Feshbach molecule
state $|\Phi_{b}^{(C)}\rangle$. With direct calculation, we find
that in the first-order processes the final states of these
transitions are $|\Psi_{+}({\bf
k})\rangle=|$$\uparrow\rangle_{1}|$$\uparrow\rangle_{2}(|{\bf
k}\rangle_{1}|-{\bf k}\rangle_{2}-|-{\bf k}\rangle_{1}|{\bf
k}\rangle_{2})/\sqrt{2}$ and $|\Psi_{-}({\bf
k})\rangle=|$$\downarrow\rangle_{1}|$$\downarrow\rangle_{2}(|{\bf
k}\rangle_{1}|-{\bf k}\rangle_{2}-|-{\bf k}\rangle_{1}|{\bf
k}\rangle_{2})/\sqrt{2}$, where $|{\bf k}\rangle_{i}$ is the
eigen-state of the momentum of atom $i$. $|\Psi_{\pm}({\bf
k})\rangle$ are eigen-states of $H_{1}$, with corresponding
eigen-energies $E_{\pm}({\bf k})=|{\bf p}|^{2}+k_{0}^{2}\mp\eta.$

Significant quantum transitions can occur between
$|\Phi_{b}^{(C)}\rangle$ and the resonant finial states
$|\Psi_{\pm}({\bf k})\rangle$, which satisfies the resonance
condition $E_{\pm}({\bf k})=E_{b}=-a^{-2}$. Since $E_{\pm}({\bf
k})\geq k_{0}^{2}\mp\eta$, when the two-photon detuning $\eta=0$,
this resonance condition cannot be satisfied by any value of ${\bf
k}$ (Fig. 3(a)). As a result, the dissociation is very weak, and the
number of remaining molecules is large. When the two-photon detuning
$\eta$ is increased so that $\eta\gtrsim k_{0}^{2}+1/a^{2}$,
$|\Phi_{b}^{(C)}\rangle$ becomes resonant with free motional states
$|\Psi_{+}({\bf k})\rangle$ (Fig. 3(b)). These states have
relatively small momentum ${\bf k}$, and thus have large overlap
with the Feshbach molecule state (i.e., large Frank-Condon factor).
Thus, the transitions from $|\Phi_{b}^{(C)}\rangle$ to these states
are significant. Similarly, when $\eta$ is tuned to be negative and
$-\eta\gtrsim k_{0}^{2}+1/a^{2}$, $|\Phi_{b}^{(C)}\rangle$ becomes
resonant with $|\Psi_{-}({\bf k})\rangle$ with small ${\bf k}$ (Fig.
3(c)) and thus the transitions to these states are strong.
Therefore, for the cases with either positive or negative $\eta$,
when the condition $|\eta|\gtrsim k_{0}^{2}+1/a^{2}$ is satisfied,
the dissociation becomes significant and the ratio $N_{{\rm
rem}}/N_{{\rm tot}}$ between the number of remaining Feshbach
molecules and half the number of all the atoms becomes very small.
In addition, when $|\eta|$ is further increased so that
$|\eta|>>k_{0}^{2}+1/a^{2}$, $|\Phi_{b}^{(C)}\rangle$ is resonant
with states $|\Psi_{\pm}({\bf k})\rangle$ with very large ${\bf k}$.
As a result, the overlap of the molecule state and $|\Psi_{\pm}({\bf
k})\rangle$ becomes very small. Thus, the dissociation rate is
decreased and $N_{{\rm rem}}/N_{{\rm tot}}$ is increased in the
region with large $|\eta|$. As shown in Fig. 2(a, c, e, g), all
these effects are observed in our experiment.

Based on the analysis above, we phenomenologically calculate the
ratio $N_{{\rm rem}}/N_{{\rm tot}}$ with Fermi's golden rule (FGR)
\cite{supp}. We perform the calculations with the same parameters as
in the experiments of Fig. 2(a, c, e, g), and illustrate our results
in Fig. 2(b, d, f, h). It is clear that our theoretical result is
qualitatively consistent with the experimental measurements.

\textit{Summary.} In summary, we experimentally and theoretically
investigate the dissociation effect of Feshbach molecules in the
presence of SO coupling. This dissociation effect is due to the
SO-coupling-induced transition from the Feshbach molecule state with
atoms in the singlet pseudo spin state to free-motion states of two
atoms in polarized pesudospin states. This work demonstrates that
SO-coupling in a Fermi gas constitutes a new dissociation tool for
Feshbach molecules.

$^{\dagger}$Corresponding author email: jzhang74@yahoo.com,
jzhang74@sxu.edu.cn
\begin{acknowledgments}
This research is supported by the National Basic Research Program of
China (Grant No. 2011CB921601, 2012CB922104), NSFC (Grant No.
11234008, 11361161002, 11222430), and Doctoral Program Foundation of
the Ministry of Education China (Grant No. 20111401130001). The
authors thank Hui Zhai and Qi Zhou for very helpful
discussions.\end{acknowledgments}

\section*{SUPPLEMENTARY MATERIAL}

\subsection*{phenomenological calculation of the number of remaining molecules}

In this supplementary material we show our phenomenological
numerical calculation based on Fermi's golden rule (FGR). According
to the FGR, for a fixed value of the Rabi frequency, the rate of
molecule-atom transition is
\begin{eqnarray}
A[\Omega,\eta]=\pi|\Omega|^{2}\int d{\bf k}\left\vert g_{+}({\bf
k})\right\vert ^{2}\delta\left(|{\bf
k}|^{2}+k_{0}^{2}-\eta+\frac{1}{a^{2}}\right)\nonumber\\
+\pi|\Omega|^{2}\int d{\bf k}\left\vert g_{-}({\bf k})\right\vert
^{2}\delta\left(|{\bf
k}|^{2}+k_{0}^{2}+\eta+\frac{1}{a^{2}}\right),\label{r-1-1}
\end{eqnarray}
where
\begin{eqnarray}
g_{\pm}({\bf k})=\frac{1}{2}\langle\Psi_{\pm}({\bf
k})|[\sigma_{x}^{(1)}+\sigma_{x}^{(2)}]|\Phi_{b}^{(C)}\rangle.
\end{eqnarray}
Since in our system the total momentum of the two aotms is
conserved, here the calculation $\langle\Psi_{\pm}({\bf
k})|[\sigma_{x}^{(1)}+\sigma_{x}^{(2)}]|\Phi_{b}^{(C)}\rangle$ is
done in the Hilbert space for the pseudospins and relative spatial
motion of the two atoms. It is pointed out that the direct
calculation gives $g_{+}({\bf k})=g_{-}({\bf k})$, and thus
$r(\eta)=r(-\eta)$. In our experiment the Rabi frequency is ramped
from $0$ to the maximum value $\Omega_{m}$ during the ramping time
$T$, i.e., we have $\Omega(t)=\Omega_{m}t/T.$ Therefore, in our
calculation we use the phenomenological equation
\begin{eqnarray}
\frac{dF(t)}{dt}=-A[\Omega(t),\eta]F(t)
\end{eqnarray}
to describe the variation of the ratio $F(t)\equiv N(t)/N_{{\rm
tm}}$ with time $t$. Here $N_{{\rm tm}}$ is the total number of the
Feshbach molecules created in our system. It is related to half the
number $N_{{\rm tot}}$ of all the atoms with the relation
\begin{eqnarray}
N_{{\rm tm}}=N_{{\rm tot}}r_P,\label{nn}
\end{eqnarray}
with $r_P$ the production rate of Feshbach molecule in our
experiment. $N(t)$ is the molecule number at time $t$, and satisfies
$N(0)=N_{{\rm tm}}$. According to this equation, after ramping Raman
beams, the fraction of remaining Feshbach molecules is
\begin{eqnarray}
\frac{N(T)}{N_{{\rm tm}}}=F(T)=\exp\left\{
-A[\Omega_{m},\eta]T/3\right\} ,\label{ft}
\end{eqnarray}
with $T$ the ramping time. Furthermore, the number $N_{{\rm rem}}$
of the remaining molecules detected in our experiment can be
expressed as $N_{{\rm rem}}=N(T)r_D$, with $r_D$ the detection rate.
Using this relation and Eq. ({\ref{nn}}), we find that the ratio
$N_{{\rm rem}}/N_{{\rm tot}}$ detected in the experiment is
\begin{eqnarray}
\frac{N_{{\rm rem}}}{N_{{\rm tot}}}=\frac{N(T)}{N_{{\rm
tm}}}r_Pr_D.\label{nr}
\end{eqnarray}
Substituting Eq. (\ref{ft}) into Eq. (\ref{nr}), we finally obtain
\begin{eqnarray}
\frac{N_{{\rm rem}}}{N_{{\rm tot}}}=\exp\left\{
-r[\Omega_{m},\eta]T/3\right\}r_Pr_D.\label{res}
\end{eqnarray}
In our calculation we take $r_Pr_D$ as a single fitting parameter,
and determine the value of this parameter by fitting the molecule
fraction in the case $\eta=0$ given by calculation and experimental
measurement.

\end{document}